\documentstyle[12pt]{article}
\begin{document}

\centerline{\Large\bf Flux Vacua Statistics for Two-Parameter Calabi-Yau's}
\begin{center}
\vskip 0.3in
A.Misra and A.Nanda\\
Indian Institute of Technology Roorkee\\
Roorkee-247 667, Uttaranchal\\
India\\
E-mail:aalokfph,aruunuec@iitr.ernet.in\\

\end{center}
\vskip 0.3 in
\begin{abstract}
We study the number of
flux vacua for type IIB string theory on an orientifold of the Calabi-Yau
expressed as a hypersurface in ${\bf WCP}^4[1,1,2,2,6]$ by evaluating a
suitable integral over the complex-structure moduli space as per the 
conjecture of Douglas and Ashok. We show that away from the singular
conifold locus, one gets a power law, and that the (neighborhood)
of the conifold locus 
indeed acts as an attractor in the (complex structure) moduli space. In the
process, we evaluate the periods near the conifold locus. We also
study (non)supersymmetric solutions near the conifold locus, and show
that supersymmetric solutions near the conifold locus do not support fluxes.
\end{abstract}

Flux compactification has become an important area of work in string/$M$-theory
enlarging the known families of allowed manifolds that would support fluxes,
from Calabi-Yau's and tori, to non-Calabi-Yau and non-K\"{a}hler manifolds
\cite{fluxes}.
The statistical counting of flux-vacua has been conjectured in \cite{AD}
to be given by a suitable integral over the complex structure moduli space.
In this short note, we consider type $IIB$ string theory compactified on
an orientifold of a compact two-parameter Calabi-Yau expressed 
as a hypersurface in ${\bf WCP}^4[1,1,2,2,6]$, and evaluate the conjectured
integral, away from and near the singular conifold locus. 
Our work, we believe, addresses for the first time, the crucial issue whether
the moduli space integral so central to the Ashok-Douglas conjecture, is
solvable in the context of compact Calabi-Yau's whose mirrors involves
{\it more than one} complex strucutre modulus. The same becomes important
given the fact that flux compactification in string theory can hope to
solve the vacuum selection problem by fixing some or all the moduli
which in turn implies fixing of {\it more than one} (around twenty)
undetermined parameters of the standard model.

We show that the
integral gives a power law for the number of allowed flux
vacua, and that the
conifold locus acts as an attractor in the complex-structure moduli space,
giving the dominant contribution to the flux-vacua counting. Surprisingly,
the result for the latter is very similar to that of the one-parameter
Calabi-Yau expressed as a hypersurface in ${\bf WCP}^4[1,1,1,1,4]$ in 
\cite{GKT}. We also investigate the compatibility of supersymmetry with
fluxes, and find that one can not have supersymmtric solutions that also
support fluxes near the conifold locus.

According to the conjecture of Ashok and Douglas, the
total number of no-scale flux vacua is given by a moduli space (${\cal M}$)
integral
$({\chi(X_4)\over24})^{b_3}\int_{\cal M}det(-R-\omega)$, 
where $X_4$ is the elliptically-fibered
Calabi-Yau four-fold in $F$-theory to which the the type $IIB$ orientifold
can be uplifted using the prescription of \cite{Sen}; $X_4$ 
for the case at hand is a Calabi-Yau hypersurface
in ${\bf WCP}^5[1,1,2,2,12,18]$, for which $\chi(X_4)=19, 728$\cite{GKTT}.

Consider the (moduli) space $FD(\tau)\times{\cal M}_{CS}$, where $FD(\tau)$
implies the
fundamental domain of the $SL(2,{\bf Z})$ group with the complex Teichmuller
space parameter
$\tau$: $|Re(\tau)|<{1\over2},\ |\tau|>1$, and ${\cal M}_{CS}$ is
the moduli space of complex structure deformations of a K\"{a}hler
manifold. We begin with the evaluation of $\int_{{\cal M}_{CS}}det(-R-\omega)$, 
${\cal M}_{CS}$
being the complex-structure moduli space with $\phi,\psi$ the relevant 
moduli and $\omega$ being the K\"{a}hler form, for $|\psi|<<1$ and around
$\phi=2$. We show that one gets a power law $r^6$ for some positive $r$. 
For points near the singular conifold locus: $\phi=1-
864\psi^6$ (one can similarly consider $\phi=-1-864\psi^6$), 
the aforementioned moduli-space integral gives a logarithmic
result $({\rm constant})1/(ln r + {\rm constant})$ when integrated
around $|\phi|<<1,\ |1-864\psi^6-\phi|<<1$.

The defining hypersurface for the Calabi-Yau is:
\begin{equation}
\label{eq:dieufCY}
x_0^2+x_1^{12}+x_2^{12}+x_3^6+x_4^6-12\psi x_0x_1x_2x_3x_4-2\phi x_1^6x_2^6=0,
\end{equation}
with $h^{1,1}=2$ and $h^{2,1}=128$. This is invariant under ${\bf Z}_2\times
{\bf Z}_6^2$, and using the Greene-Plesser construction, under this modding,
one gets the mirror manifold with $h^{1,1}=128$ and $h^{2,1}=2$. It is
thir mirror that we will be considering, or equivalently, as in \cite{GKTT},
the moduli that can appear at higher orders, are consistently set to zero.
Under the symplectic decomposition of the holomorphic three-form $\Omega$
canonical homology ($A_a, B^a, a=1,2,3$) and cohomology bases 
($\alpha_a,\beta^a$), defining the periods as $\int_{A_a}\Omega=z^a,
\int_{B^a}\Omega=F_a$, such that $\Omega=z^a\alpha_a-F_a\beta^a$. Then,
the K\"{a}hler potential $K$ is given by: $-ln(-i(\tau-{\bar\tau})-ln(-i\int_{CY}\Omega\wedge{\bar\Omega})=ln(-i(\tau-{\bar\tau}))-ln(-i\Pi^\dagger\Sigma\Pi)$,
$\Pi$ being the six-component period vector and $\Sigma=\left(\begin{array}{cc}
0& {\bf 1}_3 \\
-{\bf 1}_3 & 0 \\
\end{array}\right)$. 

In the vicinity of $\psi=0$ and $\phi$ on some regular locus,
the period vector $\Pi$ is given by:
\begin{eqnarray}
\label{eq:Pi}
& & \Pi=-(2\pi i)^3{4\sqrt{\pi}\over\Gamma^3({5\over 6})}
\left(\matrix{ -1 & 1 & 0 & 0 & 0 & 0 
\cr \frac{3}{2} & \frac{3}{2} & \frac{1}{2} & \frac{1}{2} & 
    -\left( \frac{1}{2} \right)  & -\left( \frac{1}{2} \right)
      \cr 1 & 0 & 1 & 0 & 0 & 0 \cr 1 & 0 & 0 & 0 & 0 & 0 \cr -\left( \frac{1}{2} \right)
      & 0 & \frac{1}{2} & 0 & \frac{1}{2} & 0 \cr \frac{1}{2} & \frac{1}{2} & -\left( \frac{1}
     {2} \right)  & \frac{1}{2} & -\left( \frac{1}{2} \right)  & \frac{1}{2} 
\cr}\right)\nonumber\\
& & \Biggl[u_{-{1\over6}}(\phi)\left(\begin{array}{c} 1\\0\\e^{{i\pi\over3}}\\
0\\e^{{2i\pi\over3}}\\0\\
\end{array}\right)+e^{{i\pi\over6}}u_{-{1\over6}}(-\phi)\left(\begin{array}{c}
0\\1\\0\\e^{{i\pi\over3}}\\0\\e^{{2i\pi\over3}}\\
\end{array}\right)\Biggr]
\nonumber\\
& & +(2\pi i)^3(12)^2{1\over2\pi}\psi^2
\left(\matrix{ -1 & 1 & 0 & 0 & 0 & 0 \cr \frac{3}{2} & \frac{3}{2} & \frac{1}{2} & \frac{1}{2} & 
    -\left( \frac{1}{2} \right)  & -\left( \frac{1}{2} \right)
      \cr 1 & 0 & 1 & 0 & 0 & 0 \cr 1 & 0 & 0 & 0 & 0 & 0 \cr -\left( \frac{1}{2} \right)
      & 0 & \frac{1}{2} & 0 & \frac{1}{2} & 0 \cr \frac{1}{2} & \frac{1}{2} & -\left( \frac{1}
     {2} \right)  
& \frac{1}{2} & -\left( \frac{1}{2} \right)  & \frac{1}{2} \cr  }\right)
\nonumber\\
& & .\Biggl[
u_{-{1\over2}}(\phi)\left(\begin{array}{c}1\\0\\-1\\0\\1\\0\\
\end{array}\right)+iu_{-{1\over2}}(-\phi)
\left(\begin{array}{c}
0\\1\\0\\-1\\0\\1\\
\end{array}\right)\Biggr].
\end{eqnarray}
In (\ref{eq:Pi}), 
\begin{equation}
\label{eq:udieuf}
u_\nu(\phi)=(2\phi)^\nu\ _2F_1(-{\nu\over2},-{\nu\over2}+{1\over2};
1;{1\over\phi^2}),
\end{equation}
which can be analytically continued to ${2^\nu\over\pi}\int_{-1}^1 
d\zeta{(\phi-\zeta)^\nu\over\sqrt{1-\zeta^2}}$.
We will also require:
\begin{equation}
\label{eq:dervu}
u^\prime_\nu(\phi)=2^{-1+\nu}\nu\phi^{-3+\nu}\biggl[-(\nu-1)\ _2F_1
(1-{\nu\over2},{3\over2}-{\nu\over2};2;{1\over\phi^2})+2\phi^2
\ _2F_1(-{\nu\over2},{1\over2}-{\nu\over2};1;{1\over\phi^2})\biggr].
\end{equation}
If $x\equiv\phi-2$, then using:
\begin{eqnarray}
\label{eq:uTaylorsers}
& & u_{-{1\over2}}(2)=\frac{128\,{\sqrt{6}}\,{EllipticK}(\frac{2}{3}) - 
    32\,{\sqrt{6}}\,x\,{EllipticK}(\frac{2}{3}) - 
    9\,\pi \,x\,{\ _2F_1}
(\frac{5}{4},\frac{7}{4},2,\frac{1}{4})}{384\,\pi },\nonumber\\
& & u_{-{1\over2}}(-\phi)=-i u_{-{1\over2}}(\phi);\nonumber\\
& & u_{-{1\over6}}(2)=\frac{-\left( -576\,{\ _2F_1}(\frac{1}{12},\frac{7}{12},1,\frac{1}{4}) + 
      48\,x\,{\ _2F_1}(\frac{1}{12},\frac{7}{12},1,\frac{1}{4}) + 
      7\,x\,{\ _2F_1}(\frac{13}{12},\frac{19}{12},2,\frac{1}{4}) \right) }
    {576\,2^{\frac{1}{3}}},\nonumber\\
& & u_{-{1\over6}}(-\phi)=e^{-{i\pi\over6}}u_{-{1\over6}}(\phi),
\end{eqnarray}
one gets:
\begin{eqnarray}
\label{eq:Pianalform}
& & \Pi=\nonumber\\
& & 
\{ \{ \frac{\frac{-i }{18}\,( -1 + {( -1 ) }^{\frac{1}{12}} ) \,
      {\pi }^{\frac{7}{2}}\,( -576\,
          {\ _2F_1}(\frac{1}{12},\frac{7}{12},1,\frac{1}{4}) + 
        48\,x\, {\ _2F_1}(\frac{1}{12},\frac{7}{12},1,\frac{1}{4}) + 
        7\,x\, {\ _2F_1}(\frac{13}{12},\frac{19}{12},2,\frac{1}{4}) ) }
      {2^{\frac{1}{3}}\,{ \Gamma(\frac{5}{6})}^3}\} ,\nonumber\\
& & \!\!\!\!\!\!\!\!\!\!\!\!\!\!\!\!\!\!\!\!\!\!\!
\{ \frac{-i }{72}\,\pi \,( \frac{2^{\frac{2}{3}}\,
         ( (3+2\,i)+4\,{(-1) }^{\frac{1}{12}} + 
           {(-1) }^{\frac{7}{12}} ) \,{\pi }^{\frac{5}{2}}\,
         ( -576_2F_1(\frac{1}{12},\frac{7}{12},1,\frac{1}{4}) + 
           48\,x_2F_1(\frac{1}{12},\frac{7}{12},1,\frac{1}{4}) + 
           7\,x_2F_1(\frac{13}{12},\frac{19}{12},2,\frac{1}{4}) )
           }{{ \Gamma(\frac{5}{6})}^3} \nonumber\\
& &  - 108\,\psi^2\,( -128\,{\sqrt{6}}\, {EllipticK}(\frac{2}{3}) + 
         32\,{\sqrt{6}}\,x\, {EllipticK}(\frac{2}{3}) + 
         9\,\pi \,x\, {\ _2F_1}(\frac{5}{4},\frac{7}{4},2,\frac{1}{4}) )
          ) \} ,\nonumber\\
& & \{ \frac{( \frac{1}{18} - \frac{i }{18} ) \,
      {\pi }^{\frac{7}{2}}\,( -576\,
          {\ _2F_1}(\frac{1}{12},\frac{7}{12},1,\frac{1}{4}) + 
        48\,x\, {\ _2F_1}(\frac{1}{12},\frac{7}{12},1,\frac{1}{4}) + 
        7\,x\, {\ _2F_1}(\frac{13}{12},\frac{19}{12},2,\frac{1}{4}) ) }
      {2^{\frac{1}{3}}\,{ \Gamma(\frac{5}{6})}^3}\} ,\nonumber\\
& & 
  \{ \frac{-i }{36}\,\pi \,( \frac{2^{\frac{2}{3}}\,{\pi }^{\frac{5}{2}}\,
         ( -576\, {\ _2F_1}(\frac{1}{12},\frac{7}{12},1,\frac{1}{4}) + 
           48\,x\, {\ _2F_1}(\frac{1}{12},\frac{7}{12},1,\frac{1}{4}) + 
           7\,x\, {\ _2F_1}(\frac{13}{12},\frac{19}{12},2,\frac{1}{4}) )
           }{{ \Gamma(\frac{5}{6})}^3}\nonumber\\
& & - 54\,\psi^2\,( -128\,{\sqrt{6}}\, {EllipticK}(\frac{2}{3}) + 
         32\,{\sqrt{6}}\,x\, {EllipticK}(\frac{2}{3}) + 
         9\,\pi \,x\, {\ _2F_1}(\frac{5}{4},\frac{7}{4},2,\frac{1}{4}) )
          ) \} ,\nonumber\\
& &\{ \frac{i }{72}\,\pi \,
    ( \frac{2^{\frac{2}{3}}\,{\pi }^{\frac{5}{2}}\,
         ( -576\, {\ _2F_1}(\frac{1}{12},\frac{7}{12},1,\frac{1}{4}) + 
           48\,x\, {\ _2F_1}(\frac{1}{12},\frac{7}{12},1,\frac{1}{4}) + 
           7\,x\, {\ _2F_1}(\frac{13}{12},\frac{19}{12},2,\frac{1}{4}) )
           }{{ \Gamma(\frac{5}{6})}^3}\nonumber\\ - 
& &       54\,\psi^2\,( -128\,{\sqrt{6}}\, {EllipticK}(\frac{2}{3}) + 
         32\,{\sqrt{6}}\,x\, {EllipticK}(\frac{2}{3}) + 
         9\,\pi \,x\, {\ _2F_1}(\frac{5}{4},\frac{7}{4},2,\frac{1}{4}) )
          ) \} ,\nonumber\\
& & \{ \frac{-i }{72}\,\pi \,
    ( \frac{2^{\frac{2}{3}}\,( 1 + {( -1 ) }^{\frac{7}{12}} ) \,
         {\pi }^{\frac{5}{2}}\,( -576\,
             {\ _2F_1}(\frac{1}{12},\frac{7}{12},1,\frac{1}{4}) + 
           48\,x\, {\ _2F_1}(\frac{1}{12},\frac{7}{12},1,\frac{1}{4}) + 
           7\,x\, {\ _2F_1}(\frac{13}{12},\frac{19}{12},2,\frac{1}{4}) )
           }{{ \Gamma(\frac{5}{6})}^3}\nonumber\\
& &  - 108\,\psi^2\,( -128\,{\sqrt{6}}\, {EllipticK}(\frac{2}{3}) + 
         32\,{\sqrt{6}}\,x\, {EllipticK}(\frac{2}{3}) + 
         9\,\pi \,x\, {\ _2F_1}(\frac{5}{4},\frac{7}{4},2,\frac{1}{4}) )
          ) \} \}
\nonumber\\
& & 
\end{eqnarray}
Using the numerical values of various quantities, the expression of
the K\"{a}hler potential excluding the axion-dilaton modulus, is given by:
\begin{eqnarray}
\label{eq:K/ad}
& & K=-ln(-i(\tau-{\bar\tau}) - ln\Biggl[
( 1.65278\times{10}^6 + 2.32831\times{10}^{-10}\,i  )  + 
    ( 762417 - 546469\,i  ) \,\psi^2 - \nonumber\\
& &     ( 163034 + 2.91038\times{10}^{-11}\,i  ) \,x - 
    ( 236957 - 169840\,i  ) \,\psi^2\,x + 
    ( 762417 + 546469\,i  ) \,{{\bar \psi}}^2 + 
     \nonumber\\
& &  - ( 75206.6 + 53904.9\,i ) \,x\,{{\bar \psi}}^2 - 
    ( 2.32831\times{10}^{-10}\,i ) \,|\psi|^4\,x + 
    -(163034 - 2.91038\times{10}^{-11}\,i  ) \,{\bar x}  \nonumber\\
& & - ( 75206.6 - 53904.9\,i  ) \,\psi^2\,{\bar x} + 
    ( 16082.1 - 1.81899\times{10}^{-12}\,i  ) \,|x|^2 + 
    ( 23373.9 - 16753.4\,i  ) \,\psi^2\,|x|^2\nonumber\\
& &  - ( 236957 + 169840\,i  ) \,{{\bar \psi}}^2\,
     {\bar x} + ( 2.32831\times{10}^{-10}\,i  ) \,|\psi|^4\,{\bar x} + 
    ( 23373.9 + 16753.4\,i  ) \,|x|^2\,{{\bar \psi}}^2\,\Biggr]\nonumber\\
& & \equiv -ln(-i(\tau-{\bar\tau}) + 
ln\biggl[A+(a_1 \psi^2+{\bar a}_1{\bar \psi}^2)+(b_1x+{\bar b}_1{\bar x})
+(c_1\psi^2x+{\bar c}_1{\bar \psi}^2{\bar x})
+(c_2x{\bar \psi}^2+{\bar c}_2{\bar x}\psi^2)\nonumber\\
& & +(c_3|x|^2\psi^2+{\bar c}_3|x|^2{\bar \psi}
+c_4|x|^2\biggr],\nonumber\\
& & 
\end{eqnarray}
where we drop $Im(A)\sim 10^{-10}$, $
Im(c_4)\sim 10^{-12}$ and a $c_5|\psi|^4x+{\bar c}_5|\psi|^4{\bar x}$ term with
$Re c_5=0, Im c_5\sim 10^{-12}$\footnote{Of course $ln[-i\Pi^\dagger
\Sigma\Pi]$
is real - the fact that one gets infinitesimal, but non-zero imaginary
parts as well, seems to be an artefact of using Mathematica.}.
From (\ref{eq:K/ad}), one gets the metric:
\begin{equation}
\label{eq:g}
g_{i{\bar j}}=\left(\begin{array}{ccc}
-{1\over(\tau-{\bar\tau})^2}&0&0\\
0 & 
-4{|a_1|^2\over A^2}|\psi|^2 & 2\psi({{\bar c}_2\over A} 
- {a_1{\bar b}_1\over A^2})\\
& & \\
0& 
2{\bar \psi}({{\bar c}_2\over A} - {{\bar a}_1b_1\over A^2})&-{|b_1|^2\over A^2}
+{c_4\over A}\\
\end{array}\right).
\end{equation}
One then can calculate the curvature 2-form: $R^m_{\ j{\bar k}l}dz^j\wedge
d{\bar z}^{\bar k}$ using that for a K\"{a}hler manifold, 
$R^m_{\ j{\bar k}l}={\partial\Gamma^m_{jl}\over\partial{\bar z}^{\bar k}}
={1\over2}{\partial\over\partial{\bar z}^{\bar k}}\biggl(g^{m{\bar n}}
(\partial_jg_{l{\bar n}}+\partial_lg_{j{\bar n}})\biggr)$.
So, using the metric of (\ref{eq:g}), one gets:
\begin{eqnarray}
\label{eq:;Rcomps}
& & R^m_{\ x{\bar x}l}=R^m_{\ \psi{\bar x}l}=R^m_{\ x{\bar \psi}l}=0;\nonumber\\
& & R^\psi_{\ \psi{\bar \psi}\psi}={1\over{\bar \psi}^2}{({{\bar c}_2\over A}-{a_1{\bar b}_1
\over A^2})^2\over{4[{|a_1|^2\over A^2}({|b_1|^2\over A^2}-{c_4\over A})-
|{c_2\over A}-{{\bar a}_1b_1\over A^2}|^2]}},\nonumber\\
& & R^x_{\ \psi{\bar \psi}\psi}={2|a_1|^2\over{4[{|a_1|^2\over A^2}({|b_1|^2\over A^2}-{c_4\over A})-
|{c_2\over A}-{{\bar a}_1b_1\over A^2}|^2]A^2\psi}}({{\bar c}_2\over A}
-{a_1{\bar b}_1\over A^2}),\nonumber\\
& & R^x_{\ \psi{\bar \psi}x}=R^\psi_{\ \psi{\bar \psi}x}=0,
\end{eqnarray} 
One thus gets:
\begin{equation}
\label{eq:Rfin}
R^m_l=\left(\begin{array}{ccc}
-{2\over(\tau-{\bar\tau})^2}d\tau\wedge d{\bar\tau} & 0 & 0 \\
0& {1\over{\bar \psi}^2}{({{\bar c}_2\over A} - {a_1{\bar b}_1\over A^2})\over
4[{|a_1|^2\over A^2}({|b_1|^2\over A^2}-{c_4\over A})-
|{c_2\over A}-{{\bar a}_1b_1\over A^2}|^2]}d\psi\wedge d{\bar \psi} & 0\\
&&\\
0& {2|a_1|^2({{\bar c}_2\over A} 
- {{\bar a}_1b_1\over A^2})\over 4A^2 \psi
[{|a_1|^2\over A^2}({|b_1|^2\over A^2}-{c_4\over A})-
|{c_2\over A}-{{\bar a}_1b_1\over A^2}|^2]}d\psi\wedge d{\bar \psi} & 0
\\
\end{array}\right),
\end{equation}
One can show that $R_{j{\bar k}}=\partial_j{\bar\partial}_{\bar k}
det(g_{i{\bar j}})$,
is a hermitian matrix, i.e., ${\bar R_{k{\bar j}}}=R_{j{\bar k}}$ or
equivalently $i\partial{\bar\partial}ln det(g_{i{\bar j}})$ is a hermitian
form.
However, $R^m_l\equiv R^m_{j{\bar k}l}dz^j\wedge d{\bar z}^{\bar k}$ is
not a hermitian form.\footnote{We thank D.Joyce for discussion on this point.} 
                                                                                
For the K\"{a}hler metric corresponding to the K\"{a}hler potential of
(\ref{eq:K/ad}), one gets:$R_{j{\bar k}}=\partial_j{\bar\partial_{\bar k}}
ln\biggl({|\psi|^2\over(\tau-{\bar\tau})^2}\biggr)$, i.e.,
\begin{equation}
\label{eq:Ricci2}
R_{j{\bar k}}=\left(\begin{array}{ccc}
-{1\over(\tau - {\bar\tau})^2} & 0 & 0 \\
0 & 0 & 0 \\
0 & 0 & 0 \\
\end{array}\right),
\end{equation}
is hermitian, as expected.

One has to evaluate $det(R+\omega)$ which means one has to evaluate the
determinant of:
\begin{eqnarray}
\label{eq:det}
& & 
\!\!\!\!\!\!\!\left(\begin{array}{ccc}
R^\tau_{\tau{\bar\tau}\tau}d\tau\wedge d{\bar\tau} & 0 & 0 \\
0 & 0  & 0\\
0 & 0 & 0\\
\end{array}\right)
+\left(\begin{array}{ccc}
0 & 0 & 0\\
0& 
R^\psi_{\ \psi{\bar \psi}\psi}d\psi\wedge d{\bar \psi} & R^\psi_{\ \psi{\bar \psi}x}d\psi\wedge d{\bar \psi}\\
0 & 
R^x_{\ \psi{\bar \psi}\psi}d\psi\wedge d{\bar \psi} & R^x_{\ \psi{\bar \psi}x}d\psi\wedge d{\bar \psi}\\
\end{array}\right)
+\left(\begin{array}{ccc}
0&0&0\\
0& R^\psi_{\ \psi{\bar x}\psi}d\psi\wedge d{\bar x} & R^\psi_{\ \psi{\bar x}x}d\psi\wedge d{\bar x}\\
0 & R^x_{\ \psi{\bar x}\psi}d\psi\wedge d{\bar x} & R^x_{\ \psi{\bar x}x}d\psi\wedge d{\bar x}\\
\end{array}\right)\nonumber\\
& & +\left(\begin{array}{ccc}
0 & 0 & 0 \\
0 & 
R^\psi_{\ x{\bar \psi}\psi}dx\wedge d{\bar \psi} & R^\psi_{\ x{\bar \psi}x}dx\wedge d{\bar \psi}\\
0 & 
R^x_{\ x{\bar \psi}\psi}dx\wedge d{\bar \psi} & R^x_{\ x{\bar \psi}x}dx\wedge d{\bar \psi}\\
\end{array}\right)
+\left(\begin{array}{ccc}
0 & 0 & 0 \\
0 & 
R^\psi_{\ x{\bar x}\psi}dx\wedge d{\bar x} & R^\psi_{\ x{\bar x}x}dx\wedge d{\bar x}\\
0 & 
R^x_{\ x{\bar x}\psi}dx\wedge d{\bar x} & R^x_{\ x{\bar x}x}dx\wedge d{\bar x}\\
\end{array}\right)\nonumber\\
& & 
+\left(\begin{array}{ccc}
g_{\tau{\bar\tau}}d\tau\wedge d{\bar\tau} & 0 & 0 \\
0 &g_{\tau{\bar\tau}}d\tau\wedge d{\bar\tau} & 0 \\
0 & 0 & g_{\tau{\bar\tau}}d\tau\wedge d{\bar\tau}\\
\end{array}\right)
+\left(\begin{array}{ccc}
g_{\psi{\bar \psi}}d\psi\wedge d{\bar \psi} & 0 & 0 \\
0 & g_{\psi{\bar \psi}}d\psi\wedge d{\bar \psi} & 0 \\
0 & 0 & g_{\psi{\bar \psi}}d\psi\wedge d{\bar \psi}\\
\end{array}\right)\nonumber\\
& & +\left(\begin{array}{ccc}
g_{\psi{\bar x}}d\psi\wedge d{\bar x} & 0 & 0 \\
0 & g_{\psi{\bar x}}d\psi\wedge d{\bar x} & 0 \\
0 & 0 & g_{\psi{\bar x}}d\psi\wedge d{\bar x}\\
\end{array}\right)+\left(\begin{array}{ccc}
g_{x{\bar \psi}}dx\wedge d{\bar \psi} & 0 & 0 \\
0 & g_{x{\bar \psi}}dx\wedge d{\bar \psi} & 0 \\
0 & 0 & g_{x{\bar \psi}}dx\wedge d{\bar \psi}\\
\end{array}\right)\nonumber\\
& & +\left(\begin{array}{ccc}
g_{x{\bar x}}dx\wedge d{\bar x} & 0 & 0 \\
0 & g_{x{\bar x}}dx\wedge d{\bar x} & 0 \\
0 & 0 & g_{x{\bar x}}dx\wedge d{\bar x}\\
\end{array}\right)
\end{eqnarray}
One thus gets:
\begin{eqnarray}
\label{eq:detfin}
& & \int_{FD\times{\cal M}_{CS}}det(R+\omega)=
\int_{FD}(R^\tau_{\tau{\bar\tau}\tau}+g_{\tau{\bar\tau}})d\tau\wedge d{\bar\tau}
\nonumber\\
& & \wedge\int_{{\cal M}_{CS}}
\Biggl[(R^\psi_{\ \psi{\bar \psi}\psi}+g_{\psi{\bar \psi}})(R^x_{\ x{\bar x}x}
+g_{x{\bar x}})
-(R^\psi_{\ \psi{\bar x}\psi}+g_{\psi{\bar x}})(R^x_{\ x{\bar \psi}x}
+g_{x{\bar \psi}})-
(R^\psi_{\ x{\bar \psi}\psi}+g_{x{\bar \psi}})(R^x_{\ \psi{\bar x}x}
+g_{\psi{\bar x}})\nonumber\\
& & +(R^\psi_{\ x{\bar x}\psi}+g_{x{\bar x}})(R^x_{\ \psi{\bar \psi}x}
+g_{\psi{\bar \psi}})-R^\psi_{\ \psi{\bar \psi}x}R^x_{\ x{\bar x}\psi}
+R^\psi_{\ \psi{\bar x}x}R^x_{\ x{\bar \psi}\psi}
+R^\psi_{\ x{\bar \psi}x}R^x_{\ \psi{\bar x}\psi}\nonumber\\
& & 
-R^\psi_{\ x{\bar x}x}R^x_{\ \psi{\bar \psi}\psi}\Biggr]d\psi\wedge d{\bar \psi}\wedge dx\wedge 
d{\bar x}\nonumber\\
& & 
\end{eqnarray}
One thus sees
that moduli space integral factorizes into an integral 
over the fundamental domain of $SL(2,{\bf Z})$, and an integral over the
complex structure moduli space. Given that the former will give only a
multiplicative contribution, in the following, we will concentrate only on
the complex structure moduli space integral.

Assuming that one performs the angular integrals first and then the radial
integrals, one gets:
\begin{equation}
\label{eq:detfin1}
\int det(R+\omega)\sim \int_{|x|\leq r_x;\ |\psi|\leq r_\psi}
(|\psi|^2)|x|d|x| d\theta_x |\psi| d|\psi|
d\theta_\psi\sim r_x^4 r_\psi^2\stackrel{r_x=r_\psi=L}{\rightarrow}L^6.
\end{equation}
It is interesting that the power is exactly equal to $2h^{1,1}(CY_3)+2$.

For points near the singular conifold locus: $\phi=1-864\psi^6$, the
period vector $\Pi$ is given Y:
\begin{equation}
\label{eq:conifold1}
\Pi=\left(\matrix{ -1 & 1 & 0 & 0 & 0 & 0 \cr \frac{3}{2} & \frac{3}{2} & \frac{1}{2} & \frac{1}{2} &
    -\frac{1}{2}  & - \frac{1}{2}
      \cr 1 & 0 & 1 & 0 & 0 & 0 \cr 1 & 0 & 0 & 0 & 0 & 0 \cr -\frac{1}{2} 
      & 0 & \frac{1}{2} & 0 & \frac{1}{2} & 0 \cr \frac{1}{2} & \frac{1}{2} & -\frac{1}{2}
& \frac{1}{2} & -\frac{1}{2}  & \frac{1}{2} \cr  }\right)
.\left(\begin{array}{c}
\omega_0\\ \omega_1\\ \omega_2\\ \omega_3\\ \omega_4\\ \omega_5\\
\end{array}\right),
\end{equation}
where $w_i$'s are to be determined as follows. Using \cite{Candetal,GKTT},
in the neighborhood of the conifold locus:
\begin{eqnarray}
\label{eq:percl1} 
& & w_{2j}=-{1\over6\pi^3}\sum_{r=1}^6sin^2\biggl({r\pi\over6}\biggr)sin
\biggl({r\pi\over2}\biggr)e^{2ijr\pi\over6}\xi^r_{2j},\nonumber\\
& & w_{2j+1}=-{1\over 6\pi^3}\sum_{r=1}^6sin^2\biggl({r\pi\over6}\biggr)sin
\biggl({r\pi\over2}\biggr)e^{i\pi (2j + 1)r\over6}\xi^r_{2j+1},
\end{eqnarray}
where 
\begin{eqnarray}
\label{eq:percl2}
& & \xi^r_{2j}=\sum_{n=1}^\infty{(\Gamma[n+{r\over6}])^3\Gamma[3(n+{r\over6})]
\over\Gamma[6(n+{r\over6})]}(-)^n(12\psi)^{6n+r}u_{-(n+{r\over6})}(\phi),
\nonumber\\
& & \xi^r_{2j+1}=\sum_{n=1}^\infty{(\Gamma[n+{r\over6}])^3\Gamma[3(n+{r\over6})]
\over\Gamma[6(n+{r\over6})]}(12\psi)^{6n+r}u_{-(n+{r\over6})}(-\phi).
\end{eqnarray}.
Now using the Stirling asymptotic series for the gamma function:
$\Gamma(z)\stackrel{|z|\rightarrow\infty}{\rightarrow}\sqrt{{2\pi\over z}}
\biggl({z\over e}\biggr)^z\biggl(1+{1\over12z}+{\cal O}\biggl({1\over z^2}
\biggr)\biggr)$, and the following expressions for $u_\nu(\phi)$ for
large $|\nu|$:
\begin{eqnarray}
\label{eq:percl3}
& & u_\nu(\phi)\stackrel{|\nu|\rightarrow\infty}{\rightarrow}
{2^{\nu-{1\over2}}\over\sqrt{\pi\nu}}\biggl[(1+\phi)^{\nu+{1\over2}}
-i(\phi-1)^{\nu+{1\over2}}\biggr],\nonumber\\
& & u_\nu(-\phi)\stackrel{|\nu|\rightarrow\infty}{\rightarrow}
{2^{\nu-{1\over2}}\over\sqrt{\pi\nu}}\biggl[e^{i\pi\nu}(1+\phi)^{\nu+{1\over2}}
-ie^{-i\pi\nu}(\phi-1)^{\nu+{1\over2}}\biggr],
\end{eqnarray}
one gets the following asymptotic expression for
${\partial^2w_i\over\partial\psi^2}$:
\begin{eqnarray}
\label{eq:percl4}
& & {\partial^2 w_i\over\partial\psi^2}={9\over\pi^2\psi^2}{c_i(1-\phi)\over
(1-864\psi^6-\phi)}
,\ {\rm where}\nonumber\\
& & c_{2j}\equiv{4i\over3}\sum_{r=1}^6sin^2\biggl({\pi r\over6}\biggr)sin\biggl(
{\pi r\over2}\biggr)e^{i(2j-1)\pi r\over6}(-)^r,\nonumber\\
& & c_{2j+1}\equiv{4i\over3}\sum_{r=1}^6sin^2\biggl({\pi r\over6}\biggr)
sin\biggl(
{\pi r\over2}\biggr)e^{i(2j+1)\pi r\over6}(-)^r.
\end{eqnarray}
One can show that $c_i=(1,1,-1,-2,2,1)$\cite{Candetal,GKTT}. 
\footnote{If one includes the ${1\over12z}$ in the 
Stirling expression for $\Gamma(z)$, one will get a $ln(1 - 864\psi^6 - \phi)$
term in addition to the ${1\over(1 - 864\psi^6 - \phi)}$ in 
${\partial^2 w_i\over\partial\psi^2}$.}
One thus gets:
\begin{equation}
\label{eq:percl5}
w_i={c_i\over2\pi i}\biggl({2\pi i\over 4\pi^2}{(1 - 864\psi^6 - \phi)
\over(1-\phi)^2}\biggr)ln(1 - 864\psi^6 - \phi) + f_i(\phi,\psi),
\end{equation}
where $f_i(\phi,\psi)$ are analytic functions of $\phi$ and $\psi$. Now,
the monodromy properties, as discussed in \cite{Candetal2,GKTT} imply
that:
\begin{equation}
\label{eq:percl6}
w_i={c_i\over2\pi i}(w_0 - w_1)ln(1 - 864\psi^6 - \phi) + f_i(\phi,\psi).
\end{equation}
This implies $w_0-w_1=f_0-f_1$. To determine $f_i$, we use the fact that
for $|\phi|<<1,\ |\psi|<<1$, one can expand $f_i$ as: $f_i=\sum_{n,m} a_{n,m}
\phi^n\psi^m$ \cite{Candetal2}. In the neighborhood of the conifold locus,
the analytic part of the periods $f_i(\phi,\psi)$ will be given by:
$f_0(e^{6i j\pi\phi},e^{i j\pi\over 6}\psi)$, $j=0,...,5$, where
$f_0(\phi,\psi)$ is given by ($x\equiv 1 - 864\psi^6 - \phi$):
\begin{eqnarray}
\label{eq:percl7}
& & f_0(\phi,x)=\sum_{m=1}^\infty{a_{0,m}\over(864)^m}(-)^{m\over6}
+\biggl(\sum_{m=1}^\infty{a_{1,m}\over(864)^m}(-)^{m\over6}-
\sum_{m=1}^\infty{ma_{0,m}\over(864)^m}(-)^{m\over6}\biggr)\phi\nonumber\\
& & 
-x\sum_{m=1}^\infty{m a_{0,m}\over(864)^m}
(-)^{m\over6}+{\cal O}(x^2,\phi^2,x\phi).
\end{eqnarray}
In general,
\begin{eqnarray}
\label{eq:percl8}
& & f_0(\phi,\psi)=-{1\over6}\sum_{n=1}^\infty{\Gamma({n\over6})(-12\psi)^n
u_{-{n\over6}}(\phi)\over\Gamma(n)(\Gamma[1-{n\over6}])^2
\Gamma[1-{n\over12}]}
,\nonumber\\
& & {\rm where}\nonumber\\
& & u_{-{n\over6}}(\phi)={e^{-in\pi\over12}
\over2\Gamma({n\over6})}\sum_{m=0}^\infty{e^{i\pi m\over2}
\Gamma({6m+n\over12})(2\phi)^m\over m!\Gamma(1-{6m+n\over12}
)}.
\end{eqnarray}
Now, from (\ref{eq:percl5}), one sees that $f_0-f_1\equiv\alpha x + \beta x\phi 
+ ...$, $x\equiv 1 - 864\psi^6 - \phi$, and 
\begin{equation}
\label{eq:percl9}
a_{0,m}=-{1\over12}{(-12)^me^{-i\pi m\over12}\Gamma({m\over12})\over
\Gamma(m)
(\Gamma[1-{m\over12}])^2(\Gamma[1-{m\over6}])^2}.
\end{equation}
The numerical values of the relevant infinite series are:
\begin{eqnarray}
\label{eq:percl12}
& & 12\sum_{n=1}^\infty{a_{0,n}(-)^{n\over6}\over(864)^{n\over6}}=-11.6-0.5i,
\nonumber\\
& & 12\sum_{n=1}^\infty{a_{0,n}(-)^{n\over6}n\over(864)^{n\over6}}=-1.9-1.2i,
\nonumber\\
& & 
12\sum_{n=1}^\infty{e^{i\pi n\over6}
a_{0,n}(-)^{n\over6}\over(864)^{n\over6}}=-13.3-1.4i,
\nonumber\\
& & \
12\sum_{n=1}^\infty{e^{i\pi n\over6}
a_{0,n}(-)^{n\over6}n\over(864)^{n\over6}}=-1.5+6.2i,
\nonumber\\
& & 12\sum_{n=1}^\infty{e^{i\pi n\over3}
a_{0,n}(-)^{n\over6}\over(864)^{n\over6}}=-20.5-3.5i,
\nonumber\\
& & 12\sum_{n=1}^\infty{e^{i\pi n\over3}
a_{0,n}(-)^{n\over6}n\over(864)^{n\over6}}=-12.1+24.4i,
\nonumber\\
& & 12\sum_{n=1}^\infty{e^{i\pi n\over2}
a_{0,n}(-)^{n\over6}\over(864)^{n\over6}}=-34.2-25.9i,
\nonumber\\
& & 12\sum_{n=1}^\infty{e^{i\pi n\over2}
a_{0,n}(-)^{n\over6}n\over(864)^{n\over6}}=-82.5+7.7i,
\nonumber\\
& & 12\sum_{n=1}^\infty{e^{2i\pi n\over3}
a_{0,n}(-)^{n\over6}n\over(864)^{n\over6}}=-7.1-82.5i,
\nonumber\\
& & 12\sum_{n=1}^\infty{e^{2i\pi n\over3}
a_{0,n}(-)^{n\over6}n\over(864)^{n\over6}}=-58.7-138i,
\nonumber\\
& & 12\sum_{n=1}^\infty{e^{5i\pi n\over6}
a_{0,n}(-)^{n\over6}\over(864)^{n\over6}}=81.6-50.2i,
\nonumber\\
& & 12\sum_{n=1}^\infty{e^{5i\pi n\over6}
a_{0,n}(-)^{n\over6}n\over(864)^{n\over6}}=-156.6-126.2i,
\end{eqnarray}
Similarly,
\begin{equation}
\label{eq:percl12'}
a_{1,m}=-{1\over12}{(-12)^me^{-i\pi m\over12}\Gamma({m+6\over12})\over
\Gamma(m)
(\Gamma[1-{m+6\over12}])\Gamma[1-{m\over12}](\Gamma[1-{m\over6}])^2},
\end{equation}
and the relevant numerical values of the series are:
\begin{eqnarray}
\label{eq:percl12"}
& & 
12\sum_{n=1}^\infty {a_{1,n}(-)^{n\over6}\over(864)^{n\over6}}=
-0.471-0.574i,\nonumber\\
& & 
12\sum_{n=1}^\infty {e^{in\pi\over6} e^{i\pi}
a_{1,n}(-)^{n\over6}\over(864)^{n\over6}}=
0.396 - 0.449 i,\nonumber\\
& & 12\sum_{n=1}^\infty{a_{1,n}(-)^{n\over6}e^{in\pi\over3}e^{2i\pi}
\over(864)^{n\over6}}
=-1.57 +0.972i,\nonumber\\
& & 12\sum_{n=1}^\infty{a_{1,n}(-)^{n\over6}e^{in\pi\over2}e^{3i\pi}
\over(864)^{n\over6}}
=-8.421 + 7.4362i,\nonumber\\
& & 12\sum_{n=1}^\infty{a_{1,n}(-)^{n\over6}e^{2in\pi\over3}e^{4i\pi}
\over(864)^{n\over6}}
= 15.204 + 6.422i,\nonumber\\
& & 12\sum_{n=1}^\infty{a_{1,n}(-)^{n\over6}e^{5in\pi\over6}e^{5i\pi}
\over(864)^{n\over6}}
= -18.289i.
\end{eqnarray}
One thus gets:
\begin{equation}
\label{eq:numf's}
\left(\begin{array}{c}
f_0\\f_1\\f_2\\f_3\\f_4\\f_5\\
\end{array}\right)={1\over12}\left(\begin{array}{ccc}
-11.6-0.5i & 2.811+1.626i & 1.9+1.2i \\
-13.3-1.4i & 1.896-6.649i & 1.5 - 6.2i \\
-20.5 - 3.5i & 10.53 - 2.842i & 12.1 - 24.4i \\
-34.2 - 25.9i & 7.079 - 0.264i & 8.25 - 7.7i \\
-7.1 - 82.5i & 73.904 + 144.422i & 58.7 + 138i \\
81.6-50.2i & 156.6 + 107.911i  & 156.6 + 126.2i \\
\end{array}\right)
\left(\begin{array}{c}
1\\
\phi \\
x \\
\end{array}\right)
\end{equation}
From the asymptotic analysis done earlier, writing
$f_i=a_i^{(0)}+a_i^{(1)}\phi+a_i^{(2)} x +{\cal O}(\phi^2,x^2,x\phi)$, one
knows that $f_0-f_1=(a_0^{(2)}-a_1^{(2)})x  +{\cal O}(\phi^2,x^2,x\phi)$.
From (\ref{eq:numf's}), one sees that this is approximately satisfied 
if one drops the imaginary part of $f_0$ (to an accuracy of about 95$\%$)
and also the imaginary part of $f_1$ (to an accuracy of 89.5$\%$) 
and assumes that $\phi\rightarrow0,x\rightarrow0:\frac{\phi}{x}\rightarrow0$.
One then gets, $f_0-f_1={7.4\over12} i x + {\cal O}(\phi^2,x^2,\phi x)$ (up to
$94.5\%$ accuracy).
Given that the periods can be regarded as the solutions to the homogeneous
Picard-Fuchs equation, this implies that a constant times the period vectors
will also be valid period vectors.
Consequently, one
can do a further rescaling of the periods: $w_i\rightarrow {1\over2\pi}
({12\over7.4})w_i$, to get a match with (\ref{eq:percl5}).

The K\"{a}hler potential 
is thus given by: 
\begin{eqnarray}
\label{eq:percl14}
& & K=-ln(-i(\tau - {\bar\tau}) - ln(-i\Pi^\dagger\Sigma\Pi)=\nonumber\\
& & -ln(-i(\tau-{\bar\tau}))-ln\biggl[-i\biggl({1\over2\pi}|f_0-f_1|^2ln|x|^2
+{\rm Re}({1\over2}[3{\bar f_0}+3{\bar f_1}+{\bar f_2}+{\bar f_3}
-{\bar f_4} - {\bar f_5}][-f_0+f_2+f_4] \nonumber\\
& & + [{\bar f_0}+{\bar f_2}]
[f_0+f_1-f_2+f_3-f_4+f_5]\biggr)\biggr]\nonumber\\
& & \equiv-ln(-i(\tau-{\bar\tau}) + 
\alpha|x|^2 ln|x|^2 + A + (Bx + {\bar B} {\bar x})
+(C\phi + {\bar C}{\bar\phi}) + D |x|^2 + E |\phi|^2\nonumber\\
& & + (F{\bar x}\phi + {\bar F}x {\bar\phi}) + (G\phi^2 + {\bar G}{\bar\phi}^2),
\end{eqnarray}
where $\Sigma\equiv\left(\begin{array}{cc}
0&{\bf 1}_3\\
-{\bf 1}_3 & 0 \\
\end{array}\right)$, and we have considered terms in the K\"{a}hler potential
up to quadratic in $x,\phi$ (and their complex conjugates and the products
of the same). Using $g_{i{\bar j}}=\partial_i{\bar\partial}_jK$ and from
our experience from points away from the conifold locus, we realize that 
we can exclude the axion-dilaton modulus as it will have only an overall
multiplicative effect in the moduli space integral. Thus, one gets:
\begin{eqnarray}
\label{eq:percl15}
& & g_{i{\bar j}}^{CS}=\left(\begin{array}{cc}
g_{x{\bar x}} & g_{x{\bar\phi}}\\
g_{\phi{\bar x}} & g_{\phi{\bar\phi}}
\end{array}\right)\nonumber\\
& & = \left(\begin{array}{cc}
-{\alpha\over A}ln|x|^2+{|B|^2\over A^2}-{\alpha+D\over A} &
{{\bar C}B\over A} \\
{C{\bar B}\over A} & {|C|^2\over A^2}-{E\over A}
\end{array}\right).
\end{eqnarray}
Writing
\begin{eqnarray}
\label{eq:percl16}
& & 
det g^{CS}=-{\alpha\over A}\biggl({|C|^2\over A^2} - {E\over A}\biggr)ln|x|^2+\biggl(
{|C|^2\over A^2}-{E\over A}\biggr)\biggl({|B|^2\over A^2} - {(\alpha+D)
\over A}\biggr) - {|C{\bar B}|^2\over A^2}\nonumber\\
& & \equiv\Lambda_1 ln|x|^2+\Lambda_2,
\end{eqnarray}
one gets:
\begin{eqnarray}
\label{eq:percl17}
& & R^{CS}=\left(\begin{array}{cc}
R^x_{\ x{\bar x}x}dx\wedge d{\bar x} & R^x_{\ x{\bar x}\phi}dx\wedge d{\bar x}\\
R^\phi_{\ x{\bar x}x}dx\wedge d{\bar x} & R^\phi_{\ x{\bar x}\phi}dx\wedge
d{\bar x}\\
\end{array}\right)\nonumber\\
& & ={1\over(\Lambda_1ln|x|^2+\Lambda_2)^2}\left(\begin{array}{cc}
\biggl({|C|^2\over A^2}-{E\over A}\biggr){\alpha\Lambda_1\over A|x|^2}dx\wedge 
d{\bar x} & 0 \\
-{C{\bar B}\alpha\Lambda_1\over A^2}dx\wedge d{\bar x} & 0 \\
\end{array}\right),
\end{eqnarray}
which even though is not hermitian, the corresponding $R_{i{\bar j}}=\partial_i
\partial_{\bar j}ln det g=\partial_i\partial_j ln({[\Lambda_1 ln|x|^2 + 
\Lambda_2]\over(\tau-{\bar\tau})^2})$, hence given by:
\begin{equation}
\label{eq:Riccimat}
R_{i{\bar j}}=\left(\begin{array}{ccc}
-{1\over(\tau-{\bar\tau})^2}&0&0\\
0&-{{\Lambda_1^2\over|x|^2}\over(\Lambda_1ln|x|^2+\Lambda_2)^2}&0\\
0&0&0\\
\end{array}\right),
\end{equation}
{\it is} hermitian.
Dropping the K\"{a}hler form as compared to the curvature two-form, 
as $x\rightarrow0$,
\begin{equation}
\label{eq:percl18}
\int_{\cal M}det(R+\omega)\sim\int_{|x|<r_x}\int_{|\phi|<r_\phi}|x| d|x|
|\phi|d|\phi|{{\rm Constant}\over|x|^2(ln|x|^2 + {\Lambda_2\over\Lambda_1})^2}
=-{r_\phi^2({\rm Constant})\over(ln r_x + {\Lambda_2\over2\Lambda_1})}.
\end{equation}
One thus sees, that the contribution from the conifold locus to the 
complex structure moduli space integral dominates over the contribution
from points away from the singular conifold locus, as expected. The conifold
locus, thus, acts as an attractor in the complex structure moduli space.

Lets now consider the (non)supersymmetric solutions, implying that we look
for solutions to:
\begin{equation}
\label{eq:percl19}
W=D_\tau W=D_x W=D_\phi W=0
\end{equation}
$W$ being the superpotential, given by:
\begin{equation}
\label{eq:percl20}
W=\int_{CY}(F-\tau H)\wedge\Omega=(2\pi)^2\alpha^\prime(f-\tau h)\cdot\Pi,
\end{equation}
where one uses that the $NS-NS$ flux $H$ and the $RR$ flux $F$ are given by:
$F=(2\pi)^2\alpha^\prime({\cal F}_a\beta_a+{\cal F}_{a+3}\alpha_a)$ and 
$H=(2\pi)^2\alpha^\prime({\cal H}_a\beta_a+{\cal H}_{a+3}\alpha_a)$, 
$\alpha_a,\beta^a$ 
forming an integral cohomology basis, $a=1,2,3$. For nonsupersymmetric
solutions, $W\neq0$. Now, $W=0$ together with
\begin{equation}
\label{eq:percl22}
D_\tau W=\partial_\tau W + \partial_\tau K W=-{1\over(\tau-{\bar\tau})}
({\cal F}^T\Pi-{\bar\tau}{\cal H}^T\Pi)=0,
\end{equation}
implies
\begin{equation}
\label{eq:percl23}
{\cal F}^T{\bar\Pi}={\cal H}^T{\bar\Pi}=0.
\end{equation}
The analysis below is applicable for 
nonsupersymmetric solutions, however for supersymmetric solutions, after
doing the analysis below from (\ref{eq:percl24})-(\ref{eq:percl26}), 
one notes that if in ${\cal F}^T$ and ${\cal H}^T$, one sets 
${\cal F}_4={\cal H}_4=0$, then one can write ${\cal F}^T\Pi={\cal H}^T\Pi=0$
These conditions translate to:
\begin{eqnarray}
\label{eq:W=DW=0}
& & 
-{ix\over2\pi}{\cal F}_1+{{\cal F}_2\over2}(3f_0+3f_1+f_2+f_3-f_4-f_5)
+{\cal F}_3(f_0+f_2)+{{\cal F}_5\over2}(-f_0+f_2+f_4)\nonumber\\
& & +{{\cal F}_6\over2}(f_0
+f_1-f_2+f_3-f_4+f_5)=0\nonumber\\
& & -{ix\over2\pi}{\cal H}_1+{{\cal H}_2\over2}(3f_0+3f_1+f_2+f_3-f_4-f_5)
+{\cal H}_3(f_0+f_2)+{{\cal H}_5\over2}(-f_0+f_2+f_4)\nonumber\\
& & +{{\cal H}_6\over2}(f_0
+f_1-f_2+f_3-f_4+f_5)=0,\nonumber\\
& & 
\end{eqnarray}
which implies
\begin{eqnarray}
\label{eq:solsFH1}
& & {{\cal F}_2\over2}
(3a^{(0)}_0+3a^{(0)}_1+a^{(0)}_2+a^{(0)}_3-a^{(0)}_4-a^{(0)}_5)
+{\cal F}_3(a^{(0)}_0+a^{(0)}_2)
+{{\cal F}_5\over2}(-a^{(0)}_0+a^{(0)}_2+a^{(0)}_4)\nonumber\\
& & +{{\cal F}_6\over2}(a^{(0)}_0
+a^{(0)}_1-a^{(0)}_2+a^{(0)}_3-a^{(0)}_4+a^{(0)}_5)=0\nonumber\\
& & {{\cal F}_2\over2}(3a^{(1)}_0
+3a^{(1)}_1+a^{(1)}_2+a^{(1)}_3-a^{(1)}_4-a^{(1)}_5)
+{\cal F}_3(a^{(1)}_0+a^{(1)}_2)+{{\cal F}_5\over2}(-a^{(1)}_0
+a^{(1)}_2+a^{(1)}_4)\nonumber\\
& & +{{\cal F}_6\over2}(a^{(1)}_0
+a^{(1)}_1-a^{(1)}_2+a^{(1)}_3-a^{(1)}_4+a^{(1)}_5)=0\nonumber\\
& & -{i{\cal F}_1\over2\pi} +{{\cal F}_1\over2}
(3a^{(2)}_0+3a^{(2)}_1+a^{(2)}_2+a^{(2)}_3-a^{(2)}_4-a^{(2)}_5)
+{\cal F}_3(a^{(2)}_0+a^{(2)}_2)+{{\cal F}_5\over2}
(-a^{(2)}_0+a^{(2)}_2+a^{(2)}_4)\nonumber\\
& & +{{\cal F}_6\over2}(a^{(2)}_0
+a^{(2)}_1-a^{(2)}_2+a^{(2)}_3-a^{(2)}_4+a^{(2)}_5)=0.
\end{eqnarray}
Similar equations can be written for ${\cal H}_{1,2,3,5,6}$.
It is understood that equations (\ref{eq:percl24})-(\ref{eq:percl26})
are to solved for $x,\phi$ in terms of ${\cal F}, {\cal H}$ and $\tau$.
One has to look for integer solutions (for ${\cal F}_i$ and ${\cal H}_i$)
of (\ref{eq:solsFH1}). For instance, to solve for ${\cal F}_i$, one sees
that one ends
 up with three equations in five variables. Keeping any two fixed, one
in general gets complex solutions in terms of the these fixed ${\cal F}_i$'s.
One thus has to ensure that the imaginary parts vanish and the real
parts turn out to be integral. One can show that it is not possible to
achieve the same. This imples that near the conifold locus of the
(orientifold of) the compact Calabi-Yau expressed as a degree-12 hypersurface
in ${\bf WCP}^4[1,1,2,2,6]$, one can
not obtain supersymmetric solutions that support fluxes. Note, however, 
for non supersymmetric solutions, for ${\cal H}^T\Pi\neq0,\ {\cal F}^T\Pi
\neq0$, $D_\tau W=0$ implies that $\tau={{\cal F}^T{\bar\Pi}
\over{\cal H}^T{\bar\Pi}}$.

We now consider $D_xW=0$.
\begin{eqnarray}
\label{eq:percl24} 
& & D_xW=\partial_xW+\partial_xKW=\nonumber\\
& & ({\cal F}^T - \tau {\cal H}^T)\left(\begin{array}{c}
-{i\over 2\pi}\\
{1\over2}(3a^{(2)}_0+3a^{(2)}_1+a^{(2)}_2+a^{(2)}_3-a^{(2)}_4-a^{(2)}_5\\
a^{(2)}_0+a^{(2)}_2\\
{1\over4\pi^2}(ln x + 1) + a^{(2)}_0\\
{1\over2}(-a^{(2)}_0+a^{(2)}_2+a^{(2)}_4)\\
{1\over2}(a^{(2)}_0+a^{(2)}_1-a^{(2)}_2+a^{(2)}_3-a^{(2)}_4+a^{(2)}_5)\\
\end{array}\right)\nonumber\\
& & +(4\pi({\bar x}ln|x|^2+{\bar x})+B+D{\bar x}
+{\bar F}{\bar\phi})({\cal F}^T-\tau {\cal H}^T)
\left(\begin{array}{c}
-{ix\over 2\pi}\\
{1\over2}(3f_0+3f_1+f_2+f_3-f_4-f_5)\\
f_0+f_2\\
{x\over2\pi}ln x + f_0\\
{1\over2}(-f_0+f_2+f_4)\\
{1\over2}(f_0+f_1-f_2+f_3-f_4+f_5)
\end{array}\right)\nonumber\\
& & 
=\!\!\!\!\
\frac{1}{2}\biggl(\frac{i \,( {{\cal F}_1} - {{\cal H}_1}\,\tau ) }{\pi } + 
    ( 3\,{a^{(2)}_0} + 3\,{a^{(2)}_1} + {a^{(2)}_2} + {a^{(2)}_3} - 
       {a^{(2)}_4} - {a^{(2)}_5} ) \,
     ( {{\cal F}_2} - {{\cal H}_2}\,\tau )  + 
    2\,( {a^{(2)}_0} + {a^{(2)}_2} ) \,
     ( {{\cal F}_3} - {{\cal H}_3}\,\tau )  +\nonumber\\ 
& &  ( -{a^{(2)}_0} + {a^{(2)}_2} + {a^{(2)}_4} ) \,
     ( {{\cal F}_5} - {{\cal H}_5}\,\tau ) + ({a^{(2)}_0} 
+ {a^{(2)}_1} - {a^{(2)}_2} + {a^{(2)}_3} - {a^{(2)}_4} + 
       {a^{(2)}_5} ) \,( {{\cal F}_6} - {{\cal H}_6}\,\tau )  + \nonumber\\
& & 2\,( {{\cal F}_4} - {{\cal H}_4}\,\tau ) \,
     ( {a^{(2)}_0} + \frac{1 + ln x}{4\,{\pi }^2} )  + 
    ( \frac{-i \,( {{\cal F}_1} - {{\cal H}_1}\,\tau ) }{\pi } + 
\nonumber\\
& &     
   ( 6\,{a^{(1)}_0} + {a^{(1)}_2} + {a^{(1)}_3} - {a^{(1)}_4} - 
          {a^{(1)}_5} + 3\,{a^{(2)}_0}\,x + 3\,{a^{(2)}_1}\,x + {a^{(2)}_2}\,x + 
          {a^{(2)}_3}\,x - \nonumber\\
& & {a^{(2)}_4}\,x - {a^{(2)}_5}\,x + 6\,{a^{(1)}_0}\,\phi + 
          {a^{(1)}_2}\,\phi + {a^{(1)}_3}\,\phi - {a^{(1)}_4}\,\phi - {a^{(1)}_5}\,\phi ) \
         ( {{\cal F}_2} - {{\cal H}_2}\,\tau )  + \nonumber\\
& &      
2\,( {a^{(1)}_0} + {a^{(1)}_2} + {a^{(2)}_0}\,x + {a^{(2)}_2}\,x + 
          {a^{(1)}_0}\,\phi + {a^{(1)}_2}\,\phi ) \,
        ( {{\cal F}_3} - {{\cal H}_3}\,\tau )  + \nonumber\\
& & ( -{a^{(1)}_0} + {a^{(1)}_2} + {a^{(1)}_4} - {a^{(2)}_0}\,x + 
          {a^{(2)}_2}\,x + {a^{(2)}_4}\,x - {a^{(1)}_0}\,\phi + {a^{(1)}_2}\,\phi + 
          {a^{(1)}_4}\,\phi ) \,( {{\cal F}_5} - {{\cal H}_5}\,\tau )\nonumber\\
& &  + ( 2\,{a^{(1)}_0} - {a^{(1)}_2} + {a^{(1)}_3} - {a^{(1)}_4} + 
          {a^{(1)}_5} + {a^{(2)}_0}\,x + {a^{(2)}_1}\,x - {a^{(2)}_2}\,x + 
          {a^{(2)}_3}\,x - {a^{(2)}_4}\,x + {a^{(2)}_5}\,x + 2\,{a^{(1)}_0}\,\phi 
\nonumber\\
& & 
- {a^{(1)}_2}\,\phi + {a^{(1)}_3}\,\phi - {a^{(1)}_4}\,\phi + {a^{(1)}_5}\,\phi ) \
( {{\cal F}_6} - {{\cal H}_6}\,\tau )\nonumber\\
& &   + 2\,( {{\cal F}_4} - {{\cal H}_4}\,\tau ) \,
        ( {a^{(1)}_0} + {a^{(2)}_0}\,x + {a^{(1)}_0}\,\phi + 
          \frac{x\, ln x}{4\,{\pi }^2} )  ) \,
     ( B + 2\,H,x + {\bar F}\,{\bar \phi} + 
       {\bar x}\,({D} + 4\,\pi  + 
          4\,\pi \,ln |x|^2 )  )\biggr)=0\nonumber\\
& & 
\end{eqnarray}
Near $x=0,\ \phi=0$, and for nonsupersymmetric solutions (for which
${\cal F}_4\neq0,\ {\cal H}_4\neq0$) one gets
\begin{eqnarray}
\label{eq:percl25}
& & ln x = \frac{4\,{\pi }^2}{{{\cal F}_4} - {{\cal H}_4}\,\tau}
\,\biggl( \frac{\frac{-i }{2}\,
         ( {{\cal F}_1} - {{\cal H}_1}\,\tau ) }{\pi } - 
      \frac{( 3\,{a^{(2)}_0} + 3\,{a^{(2)}_1} + {a^{(2)}_2} + {a^{(2)}_3} - 
           {a^{(2)}_4} - {a^{(2)}_5} ) \,
         ( {{\cal F}_2} - {{\cal H}_2}\,\tau ) }{2} - 
\nonumber\\      
& & ( {a^{(2)}_0} + {a^{(2)}_2} ) \,
       ( {{\cal F}_3} - {{\cal H}_3}\,\tau )  - 
      \frac{( -{a^{(2)}_0} + {a^{(2)}_2} + {a^{(2)}_4} ) \,
         ( {{\cal F}_5} - {{\cal H}_5}\,\tau ) }{2} - \nonumber\\
& & \frac{( {a^{(2)}_0} + {a^{(2)}_1} - {a^{(2)}_2} + {a^{(2)}_3} - 
           {a^{(2)}_4} + {a^{(2)}_5} ) \,
         ( {{\cal F}_6} - {{\cal H}_6}\,\tau ) }{2} - \nonumber\\
& &       \frac{B}{2}\,( \frac{-i \,( {{\cal F}_1} - {{\cal H}_1}\,z ) }{\pi } + 
           ( 6\,{a^{(0)}_0} + {a^{(0)}_2} + {a^{(0)}_3} - {a^{(0)}_4} - 
              {a^{(0)}_5} ) \,( {{\cal F}_2} - {{\cal H}_2}\,\tau )  + \nonumber\\
& &            2\,( {a^{(0)}_0} + {a^{(0)}_2} ) \,
            ( {{\cal F}_3} - {{\cal H}_3}\,\tau )  + 
           2\,{a^{(0)}_0}\,( {{\cal F}_4} - {{\cal H}_4}\,\tau )  + 
           ( -{a^{(0)}_0} + {a^{(0)}_2} + {a^{(0)}_4} ) \,
            ( {{\cal F}_5} - {{\cal H}_5}\,\tau )  + \nonumber\\
& &    ( 2\,{a^{(0)}_0} - {a^{(0)}_2} + {a^{(0)}_3} - {a^{(0)}_4} + 
        {a^{(0)}_5} ) \,( {{\cal F}_6} - {{\cal H}_6}\,\tau )  )
          \biggr) 
\end{eqnarray}
For nonsupersymmetric solutions, for which ${\cal F}_4={\cal H}_4=0$, 
one gets a linear equation in $\phi$ and $x$ from (\ref{eq:percl24}); there
will be no $ln x$ or $ln{\bar x}$ terms.

Lets now consider $D_\phi W=0$:
\begin{eqnarray}
\label{eq:percl26}
& & D_\phi W=\partial_\phi W + \partial_\phi KW\nonumber\\
& & =({\cal F}^T - \tau {\cal H}^T)\left(\begin{array}{c}
0 \\
{1\over2}(3a_0^{(1)}+3a^{(1)}_1+a^{(1)}_2+a^{(1)}_3-a^{(1)}_4-a^{(1)}_5)\\
a^{(1)}_0+a^{(1)}_2 \\
a^{(1)}_1\\
{1\over2}(-a^{(1)}_0+a^{(1)}_2+a^{(1)}_4\\
{1\over2}(a^{(1)}_0+a^{(1)}_1-a^{(1)}_2+a^{(1)}_3-a^{(1)}_4+a^{(1)}_5\\
\end{array}\right)
\nonumber\\
& & +(C + E{\bar\phi} + F{\bar x}+2G\phi)({\cal F}^T-\tau {\cal H}^T)
\left(\begin{array}{c}
-{ix\over2\pi} \\
{1\over2}(3f_0+3f_1+f_2+f_3-f_4-f_5)\\
{x\over4\pi^2}ln x +f_0\\
{1\over2}(-f_0+f_2+f_4)\\
{1\over2}(f_0+f_1-f_2+f_3-f_4+f_5)\\
\end{array}\right)\nonumber\\
& & =\frac{1}{2}\biggl(( 
3\, {a^{(1)}_0} + 3\,{a^{(1)}_1} + {a^{(1)}_2} + {a^{(1)}_3} - 
       {a^{(1)}_4} - {a^{(1)}_5} ) \,
     ( {{\cal F}_2} - {{\cal H}_2}\,\tau )  + 
    2\,( {a^{(1)}_0} + {a^{(1)}_2} ) \,
     ( {{\cal F}_3} - {{\cal H}_3}\,\tau )  + \nonumber\\
& &\!\!\!\!\!\!\!\!\!\     2\,{a^{(1)}_0}\,( {{\cal F}_4} - {{\cal H}_4}\,\tau )  + 
    ( -{a^{(1)}_0} + {a^{(1)}_2} + {a^{(1)}_4} ) \,
     ( {{\cal F}_5} - {{\cal H}_5}\,\tau )  + 
    ( {a^{(1)}_0} + {a^{(1)}_1} - {a^{(1)}_2} + {a^{(1)}_3} - {a^{(1)}_4} + 
       {a^{(2)}_5} ) \,( {{\cal F}_6} - {{\cal H}_6}\,\tau )  + \nonumber\\
& & (C + 2\,G\,\phi + F\,{\bar x} + e\, {\bar\phi} ) \,
     ( \frac{-i \,( {{\cal F}_1} - {{\cal H}_1}\,\tau ) }{\pi } + 
       ( 6\,{a^{(0)}_0} + {a^{(0)}_2} + {a^{(0}_3} - {a^{(0)}_4} - 
    {a^{(0)}_5} +\nonumber\\
& &  3\,{a^{(2)}_0}\,x + 3\,{a^{(2)}_1}\,x + {a^{(2)}_2}\,x + 
    {a^{(2)}_3}\,x - {a^{(2)}_4}\,x - {a^{(2)}_5}\,x + 6\,{a^{(1)}_0}\,\phi + 
\nonumber\\    
& & {a^{(1)}_2}\,\phi + {a^{(1)}_3}\,\phi - {a^{(1)}_4}\,\phi - {a^{(1)}_5}\,
\phi ) \
         ,( {{\cal F}_2} - {{\cal H}_2}\,\tau )  + 
       2\,( {a^{(0}_0} + {a^{(0}_2} + {a^{(2)}_0}\,x + {a^{(2)}_2}\,x + 
\nonumber\\ 
& &          {a^{(1)}_0}\,\phi + {a^{(1)}_2}\,\phi ) \,
        ( {{\cal F}_3} - {{\cal H}_3}\,\tau )  + 
      ( -{a^{(0)}_0} + {a^{(0)}_2} + {a^{(0)}_4} - {a^{(2)}_0}\,x + 
          {a^{(2)}_2}\,x + {a^{(2)}_4}\,x - {a^{(1)}_0}\,\phi + {a^{(1)}_2}\,
\phi + 
\nonumber\\    
& &       {a^{(1)}_4}\,\phi ) \,( {{\cal F}_5} - {{\cal H}_5}\,\tau )  + 
       ( 2\,{a^{(0)}_0} - {a^{(0)}_2} + {a^{(0)}_3} - {a^{(0)}_4} + \nonumber\\
& &  {a^{(0)}_5} + {a^{(2)}_0}\,x + {a^{(2)}_1}\,x - {a^{(2)}_2}\,x + 
          {a^{(2)}_3}\,x - {a^{(2)}_4}\,x + {a^{(2)}_5}\,x + 2\,{a^{(1)}_0}\,
\phi-
\nonumber\\ 
& & \!\!\!\!\!\!\!\!\!\   
{a^{(1)}_2}\,\phi + {a^{(1)}_3}\,\phi - {a^{(1)}_4}\,\phi 
+ {a^{(1)}_5}\,\phi ) \
        ( {{\cal F}_6} - {{\cal H}_6}\,\tau )  +   2\,( {{\cal F}_4} - {{\cal H}_4}\,\tau ) \,
        ( {a^{(0)}_0} + {a^{(2)}_0}\,x + {a^{(1)}_0}\,\phi + 
          \frac{x\, ln x}{4\,{\pi }^2} )  )\biggr)=0
\end{eqnarray}
Substituting 
(\ref{eq:percl25}) in (\ref{eq:percl26}), one gets a single complex
constraint on $\phi$ and ${\bar\phi}$. 

The fact that we are able to show that the conifold locus acts like
an attractor in the complex structure moduli space is what can also be
argued qualitatively given that the moduli space integral involves the
curvature tensor and hence the contribution to the same will be dominated
from locii where the curvature tensor diverges.
As part of future work, one could try to solve equations (\ref{eq:percl24})
- (\ref{eq:percl26}) numerically and perform a Monte-Carlo simulation
like \cite{GKT} to verify, e.g., that the conifold locus indeed involves
clustering of solutions, and hence acts like an attractor in the moduli space..

After completion of this work, we became aware of \cite{CQ}, which has
overlap with our work.

\section*{Acknowledgement}

One of us (AM) would like to thank S.Kachru and specially A.Giryavets and
D.Joyce for several useful clarifications.
We thank J.Conlon for pointing out a typo in 
the earlier version as well as a useful communication.

\end{document}